\documentclass[letterpaper]{article} 
\usepackage{aaai2027}  
\usepackage[hyphens]{url}  
\usepackage{graphicx} 
\usepackage{natbib}  
\usepackage{caption} 
\usepackage{amsmath}
\usepackage{amssymb}
\usepackage{algorithm}
\usepackage{algorithmic}
\usepackage{booktabs}
\usepackage{array}
\usepackage{multirow}
\usepackage{etoolbox}
\makeatletter
\patchcmd{\@maketitle}
  {\footnote{Corresponding author\ifaaai@corrmulti{s}\fi.}}
  {\footnote{Corresponding author\ifaaai@corrmulti{s}\fi.\\
  \hbox to 1.8em{\hss\textsuperscript{\ensuremath{\ddagger}}}Work done during an internship at Kuaishou Technology.}}
  {}
  {\PackageWarning{author-note}{Could not patch the corresponding-author footnote}}
\makeatother
\title{
Once Generated, Ranked: End-to-End Generative Slate Recommendation with Unified Semantic-Collaborative IDs}
\author{
Yang Hu\equalcontrib\textsuperscript{\rm 1,2,\ensuremath{\ddagger}},
Jiayi Guo\equalcontrib\textsuperscript{\rm 1,3,\ensuremath{\ddagger}},
Jingui Ma\textsuperscript{\rm 2},
Ning Li\corresponding\textsuperscript{\rm 1},
Jiangling Qin\textsuperscript{\rm 1},
Yanming Li\textsuperscript{\rm 1},
Yang Deng\textsuperscript{\rm 2},
Xiaoshuang Chen\corresponding\textsuperscript{\rm 1},
Kaiqiao Zhan\textsuperscript{\rm 1}
}
\affiliations{
\textsuperscript{\rm 1}Kuaishou Technology\\
\textsuperscript{\rm 2}Peking University\\
\textsuperscript{\rm 3}Nanjing University\\
yanghu@stu.pku.edu.cn, jiayiguo@smail.nju.edu.cn
}

\begin{document}

\maketitle

\begin{abstract}
Slate Recommendation treats a slate rather than an individual item as the fundamental recommendation unit, requiring list-wise optimization of item interactions and overall slate utility. However, existing approaches rely on cascaded pipelines that separate candidate generation from ranking and constrain optimization within retrieved candidates. Recently, Generative Recommendation based on Semantic ID (SID) has provided a promising direction toward end-to-end recommendation. However, existing SID construction methods often lack recommendation-aware semantics and effective integration of local collaborative signals, while current generative recommenders optimize with next-token prediction objectives that are misaligned with slate-level optimization and overall utility. To address these challenges, we propose OGR, an end-to-end generative slate recommendation framework that directly generates ordered slates without separating generation and ranking---``\textbf{Once Generated, Ranked}.'' First, we introduce TUSID, a recommendation-aware SID construction framework that adaptively integrates item-specific semantic information and local collaborative signals to construct hierarchical SIDs for generative slate recommendation. Based on these SIDs, OGR performs list-wise preference planning to model global slate preferences and inter-item dependencies, enabling pipelined position-wise SID decoding to directly generate ordered slates. Furthermore, we propose SPA, which leverages reward-guided conservative policy optimization to align generated slates with user preferences beyond likelihood imitation. Extensive offline experiments demonstrate the effectiveness of OGR over representative baselines (\textbf{+48.2\%/+27.2\%} relative gains in NDCG@5 on industrial/public datasets), while online A/B testing on Kuaishou, a large-scale short-video platform, further confirms its practical value, yielding a \textbf{1.120\%} improvement in Effective Views.
\end{abstract}


\section{Introduction}
Slate Recommendation treats a slate rather than an individual item as the fundamental recommendation unit, requiring joint modeling of item interactions and optimization of list-wise objectives.
Existing approaches typically follow a cascaded paradigm of candidate generation–ranking–reranking, where candidate pools are constructed through multi-stage retrieval and filtering.
Although efficient for practical deployment, its optimization scope is constrained by the candidate space, making it impossible to recover potentially relevant items that are not retrieved.
Moreover, the objective misalignment across different stages of the cascaded pipeline prevents end-to-end optimization, leaving candidate generation and final ranking as separated processes---``\textbf{Generated, Then Ranked}.'' 

Recently, Generative Recommendation (GR) \cite{TIGER,OneRec} based on SID has emerged as a promising direction to overcome the limitations of candidate-constrained recommendation.
By mapping items into discrete SID spaces, generative recommenders can directly generate target items from user histories, enabling an end-to-end recommendation paradigm.
However, existing SID construction methods mainly focus on semantic discretization and inject collaborative information through representation alignment or contrastive regularization~\cite{QARM,LETTER,QuaSID}, but do not explicitly encode distance-weighted local co-occurrence distributions for slate construction.
Moreover, existing GR methods usually optimize autoregressive SID generation with the Next Token Prediction (NTP) objective, which is inherently misaligned with the list-wise optimization of overall slate utility.
Additionally, such token-level likelihood optimization focuses on predicting individual SID tokens rather than capturing user preference over the entire slate, limiting the potential of GR in Slate Recommendation.

To address these challenges, we first propose Two-stage Unified SID Construction (TUSID), tailored to generative slate recommendation.
TUSID extracts rich item information from multimodal semantics and structured high-level attributes, and performs recommendation-aware adaptive fusion to obtain semantic representations that are customized to different items and recommendation scenarios.
Furthermore, we introduce a CountSketch-based collaborative injection mechanism to incorporate local user behavioral co-occurrence structures, together with confidence-aware fusion to regulate the contribution of collaborative signals.
The resulting unified representations are quantized by residual K-means (RQ-KMeans)~\cite{OneRec} into recommendation-aware hierarchical SIDs.

Based on the proposed SID, we further introduce OGR, an end-to-end generative slate recommendation framework.
Different from existing generative methods that focus on next-item generation, OGR uses a listwise preference planner to capture the global slate structure and dependencies across display positions.
Once the planned preference representation for a position becomes available, its SID chain is decoded while the planner proceeds to subsequent positions, forming a pipeline that avoids serial item-by-item SID generation.
We further introduce Slate-Level Preference Alignment (SPA), which derives calibrated preference signals from user-feedback-based primary rewards and slate-level auxiliary rewards.
Conservative policy optimization then aligns generated slates beyond likelihood imitation while preventing excessive deviation from the reference policy.
Together, OGR unifies item-level generation and listwise organization---``\textbf{Once Generated, Ranked}.''

Extensive experiments on both large-scale industrial data and public benchmarks demonstrate the effectiveness of OGR, while online A/B testing on Kuaishou, a real-world recommendation platform serving hundreds of millions of daily active users, further confirms its practical value.

Our main contributions are summarized as follows:
\begin{itemize}
    \item We propose TUSID, a recommendation-aware SID construction framework that adaptively integrates item-specific semantic information and local collaborative signals to construct hierarchical SIDs for generative slate recommendation.
    \item We propose OGR, an end-to-end generative slate recommendation framework that combines list-wise preference planning with pipelined position-wise SID decoding. We further introduce SPA to align generated slates with user preferences through calibrated multi-objective rewards and conservative policy optimization. We further introduce SPA to align generated slates with user preferences through calibrated multi-objective rewards and conservative policy optimization.
    \item Extensive offline experiments on industrial and public datasets, together with online A/B testing results, demonstrate the effectiveness and practical impact of our methods.
\end{itemize}

\begin{figure*}[t]
\centering
\includegraphics[width=0.95\textwidth]{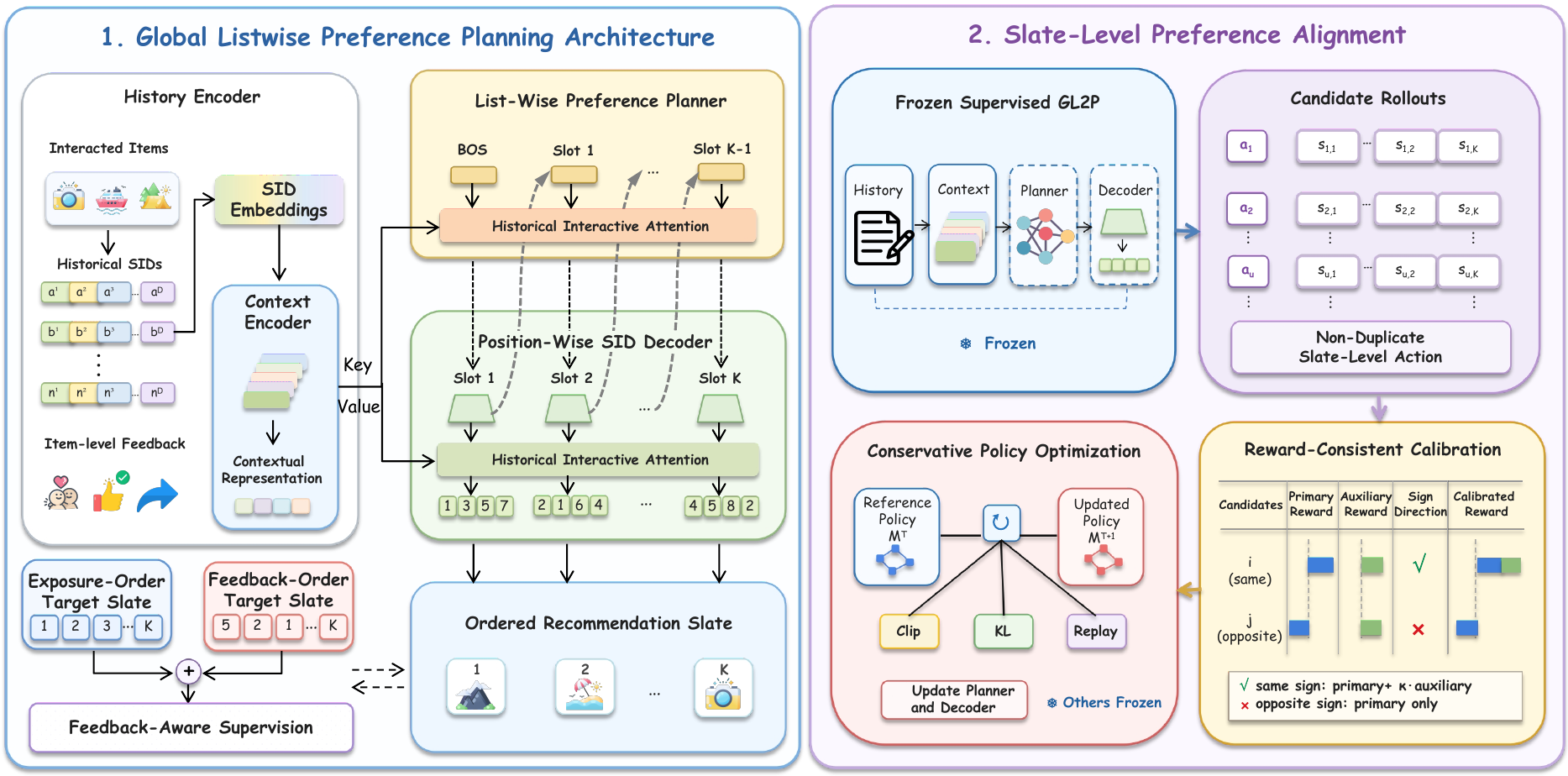} 
\caption{The overall framework of OGR.}
\vspace{-3mm}
\label{fig:1}
\end{figure*}

\section{Related Works}

Slate Recommendation jointly selects and orders multiple items by modeling inter-item interactions and listwise utility.
However, existing methods~\cite{PRM,Seq2Slate,NAR4Rec,MPAD,SAR} are typically formulated over retrieved candidate sets or predefined action spaces, leaving slate quality bounded by upstream selection.
Pioneered by TIGER~\cite{TIGER}, SID-based GR enables end-to-end recommendation by generating hierarchical semantic identifiers instead of atomic item IDs in conventional sequential recommenders~\cite{CL4SRec,DuoRec,DCRec,FEARec,BSARec,LRURec}.
Subsequent studies extend this paradigm through recommendation-aware SID construction with semantic--collaborative alignment~\cite{LCRec,LETTER,CoST,ColaRec,SemanticConvergence,UNGER}, joint SID--recommendation optimization~\cite{STORE,ETEGRec}, and efficient
generation through tree-structured or hybrid retrieval, order-agnostic and parallel generation, and decoding/cache acceleration~\cite{SEATER,LIGER,RPG,TokenRec,SETRec,AtSpeed,EARN}.
However, prevailing NTP objectives optimize the likelihood of individual SID tokens rather than inter-item dependencies and overall slate utility, limiting the effectiveness of the generative paradigm for Slate Recommendation.

\section{Methodology}


As illustrated in Figure~\ref{fig:1}, we propose \textsc{OGR}, an end-to-end generative slate recommendation framework that directly generates ordered slates conditioned on user histories, comprising TUSID for recommendation-aware SID construction, GL2P for list-wise slate generation, and SPA for preference alignment.

\subsection{Problem Formulation}

For a user \(u\), let \(\mathcal{H}_u=(i_1,\ldots,i_T)\) denote the chronological interaction history, where \(i_t\) denotes the item at the \(t\)-th interaction.
Each item \(i\) is represented by a hierarchical \(D\)-level SID \(\mathbf{s}_i=(s_i^1,\ldots,s_i^D)\).
Given \(\mathcal{H}_u\), OGR parameterized by \(\theta\) directly generates an ordered SID slate \(\widehat{\mathcal{S}}_u\) of size \(K\) and maps it into an item slate:
\begin{equation}
\widehat{\mathcal{S}}_u=F_\theta(\mathcal{H}_u),
\qquad
\widehat{\mathcal{Y}}_u=\mathrm{Map}(\widehat{\mathcal{S}}_u),
\label{eq:task}
\end{equation}
where \(\mathrm{Map}(\cdot)\) denotes SID-to-item lookup.
Training uses the observed exposure slate \(\mathcal{Y}_u=(y_1,\ldots,y_K)\) and associated feedback signals \(\mathcal{W}_u=(w_1,\ldots,w_K)\), where \(y_k\) and \(w_k\) denote the item and its feedback at slate position \(k\), respectively.

\begin{figure*}[t]
\centering
\includegraphics[width=0.95\textwidth]{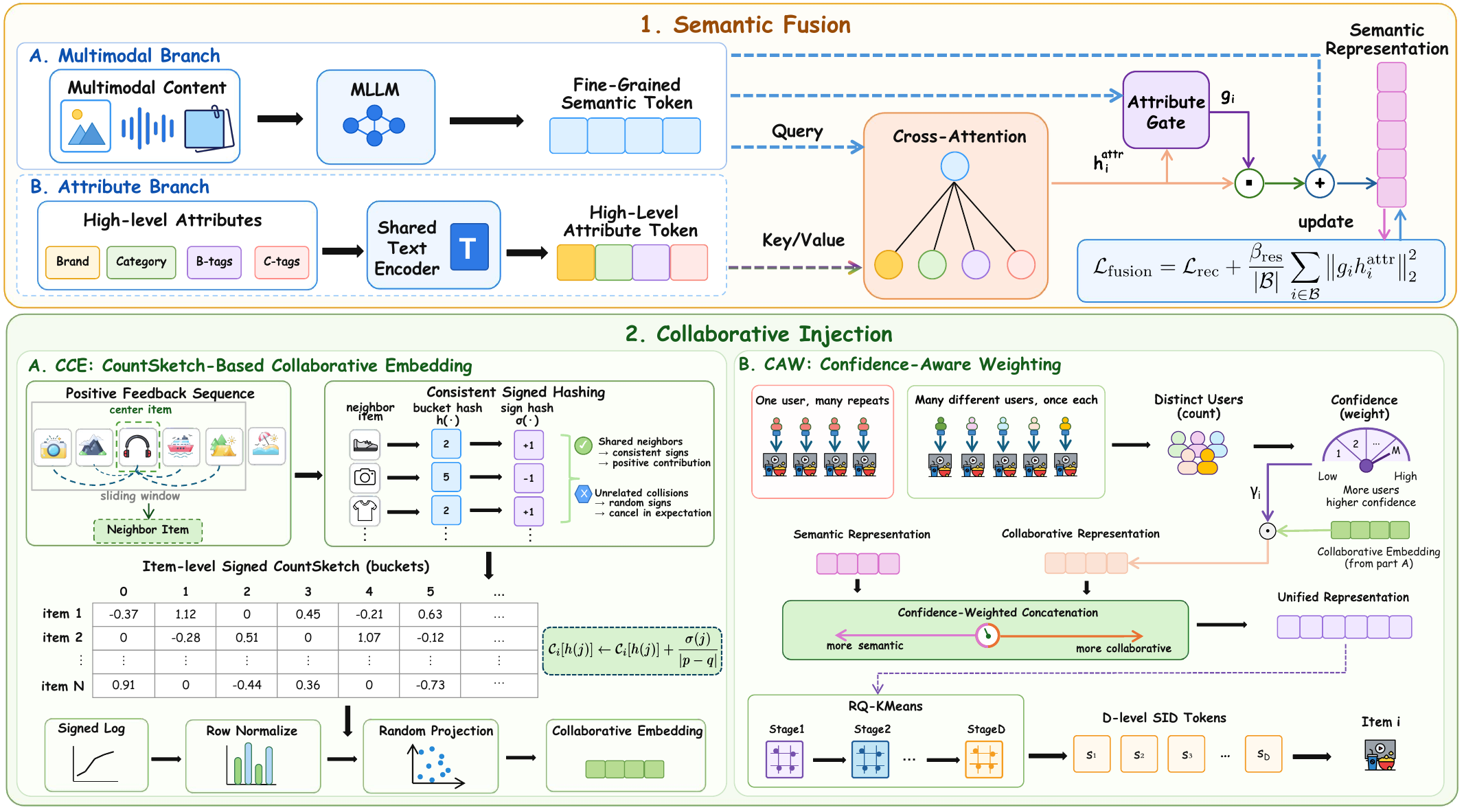} 
\caption{The overall framework of TUSID.}
\vspace{-4mm}
\label{fig:2}
\end{figure*}

\subsection{Two-Stage Unified SID Construction (TUSID)}

To construct SIDs tailored to generative slate recommendation, we propose TUSID, as illustrated in Figure~\ref{fig:2}.


\subsubsection{Semantic Fusion}

Existing SID methods mainly rely on general content representations, which may underrepresent structured high-level attributes whose relevance varies across items in recommendation scenarios.
For example, products may rely more on brand-related attributes, whereas explainer videos may benefit more from style- or audience-related attributes.
To obtain item-specific semantic representations, we encode each item from two complementary views.
An MLLM encodes the image, audio, and text of item \(i\) into a fine-grained semantic representation \(e_i^{\mathrm{M}}\), while a shared text encoder produces four high-level attribute representations \(e_i^{a}\), where \({\mathrm{a}}\in\{\mathrm{b},\mathrm{c},\mathrm{Bt},\mathrm{Ct}\}\) denotes Brand, Category, B-tags (supply-side content characteristics), and C-tags (demand-side user value), respectively.
The complete attribute taxonomy is provided in the appendix.

Since different items may rely on different high-level attributes, we retain the MLLM representation as the semantic backbone and use it to adaptively aggregate item-specific attribute information.
Specifically, \(e_i^{\mathrm{M}}\) serves as the query \(Q_i\), while \(\mathcal{E}_i^{\mathrm{attr}}=[ e_i^{\mathrm{b}}, e_i^{\mathrm{c}}, e_i^{\mathrm{Bt}}, e_i^{\mathrm{Ct}}]^\top\) provides the keys \(K_i\) and values \(V_i\).
The aggregated attribute representation is computed as
\begin{equation}
h_i^{\mathrm{attr}}
=
\operatorname{CrossAttention}
\left(
Q_i=e_i^{\mathrm{M}},
K_i=V_i=\mathcal{E}_i^{\mathrm{attr}}
\right).
\end{equation}


To regulate the overall attribute contribution, we predict a scalar sigmoid gate \(g_i\) from \([e_i^{\mathrm{M}};h_i^{\mathrm{attr}}]\) and perform gated residual fusion:

\begin{equation}
e_i^{\mathrm{sem}}
=
 e_i^{\mathrm{M}}+g_i h_i^{\mathrm{attr}}.
\label{eq:gated attr}
\end{equation}

This asymmetric fusion preserves the MLLM representation as the semantic backbone while introducing item-adaptive attribute residuals.

To make the fused representations recommendation-aware, we use \(e_i^{\mathrm{sem}}\) as the item embeddings in SASRec~\cite{SASRec} and train the fusion parameters with its next-item prediction loss \(\mathcal{L}_{\mathrm{rec}}\).
We further regularize the attribute residual to prevent attribute branch from overwhelming the MLLM backbone:
\begin{equation}
\mathcal{L}_{\mathrm{fusion}}
=
\mathcal{L}_{\mathrm{rec}}
+\frac{\beta_{\mathrm{res}}}{|\mathcal{B}|}
\sum_{i\in\mathcal{B}}
\left\|
g_i h_i^{\mathrm{attr}}
\right\|_2^2,\\
\label{eq:fusion-objective}
\end{equation}
where \(\mathcal{L}_{\mathrm{fusion}}\) denotes the overall fusion objective, \(\mathcal{B}\) denotes the set of positive and sampled negative items in SASRec, and \(\beta_{\mathrm{res}}\) controls the residual regularization strength.

\subsubsection{Collaborative Injection}

Although Semantic Fusion aligns semantic representations with recommendation objectives through \(\mathcal{L}_{\mathrm{rec}}\), it fails to capture local item-item co-occurrence structures.
We instead inject collaborative information before quantization, treating collaboration as an evidence-weighted complement, allowing items with sufficient behavioral evidence to exploit collaborative structures while preserving original semantic information for sparse and cold-start items.

\noindent\textbf{CountSketch-Based Collaborative Embedding (CCE).}

For each user \(u\), we construct co-occurrences only from the high-quality positive-feedback training prefix, excluding validation and test interactions to prevent information leakage.
For a center item \(i\) at position \(p\), each item \(j\) at position \(q\) satisfying \(0<|p-q|\leq \omega\) is treated as a contextual item and assigned the distance-based weight \(1/|p-q|\) since nearby interactions may indicate stronger local interest correlations.

To avoid storing the full item-item co-occurrence matrix, we employ Signed CountSketch~\cite{CountSketch} to compress the local contextual distribution of each item into a fixed-dimensional statistical vector. For a center item $i$ and its contextual item $j$, the update process is defined as:
\begin{equation}
\mathcal{C}_i[h(j)]
\leftarrow
\mathcal{C}_i[h(j)]
+
\frac{\sigma(j)}{|p-q|},
\label{eq:countsketch}
\end{equation}
where \(\mathcal{C}_i\) is the collaborative sketch of item \(i\), \(h(j)\) maps contextual item \(j\) to a sketch coordinate, and \(\sigma(j)\in\{-1,+1\}\) is its sign hash.
Because \(h(\cdot)\) and \(\sigma(\cdot)\) are shared across sketches, a common contextual item contributes positively to their inner product, whereas cross terms caused by unrelated collisions have zero expectation under the random sign hash.

We further apply signed-log transformation to reduce the dominance of frequent co-occurrences, followed by normalization and random projection to obtain the dense collaborative embedding \(e_i^{\mathrm{col}}\) with the same dimensionality as \(e_i^{\mathrm{sem}}\).

\noindent\textbf{Confidence-Aware Weighting (CAW).}

During CountSketch construction, we record \(U_i\), the number of distinct users contributing valid contextual updates to item \(i\), as a measure of behavioral support while limiting the influence of repeated interactions from highly active users.

Based on \(U_i\), we define the collaborative confidence \(\gamma_i\) and fusion weight \(\alpha_i\) as:
\begin{equation}
\gamma_i=
\frac{\log(1+U_i)}
{\log(1+U_i)+\tau},
\qquad
\alpha_i=\alpha_{\mathrm{col}}\gamma_i,
\end{equation}
where \(\tau>0\) controls the saturation speed of the confidence and \(\alpha_{\mathrm{col}}\in[0,1]\) is the maximum collaborative weight.

A larger \(U_i\) increases the contribution of collaborative information, whereas a smaller \(U_i\) downweights unreliable behavioral evidence for interaction-sparse or cold-start items, and when \(U_i=\gamma_i=\alpha_i=0\), the representation relies entirely on the semantic branch, consistent with the common industrial cold-start practice of using content information to compensate for sparse behavioral evidence.

After \(\ell_2\) normalization, we construct \(e_i^{\mathrm{uni}}\) by confidence-weighted concatenation:
\begin{equation}
e_i^{\mathrm{uni}}
=
\left[
\sqrt{1-\alpha_i}e_i^{\mathrm{sem}};
\sqrt{\alpha_i}e_i^{\mathrm{col}}
\right].
\label{eq:semantic-collaborative-fusion}
\end{equation}

The square-root coefficients preserve its norm, after which RQ-KMeans quantizes \(e_i^{\mathrm{uni}}\) into the final SIDs.

\subsection{Global Listwise Preference Planning Architecture (GL2P)}

Existing GR methods usually optimize autoregressive SID generation with NTP loss, which is misaligned with the slate-level optimization of item selection and ordering, limiting the explicit modeling of global slate-level preferences and inter-item dependencies.
Different from next-item generation, \textsc{OGR} treats the entire slate as the fundamental prediction unit by combining listwise preference planning with pipelined position-wise SID decoding.

\subsubsection{History Encoder}

The history encoder $\mathrm{Enc}_{\mathrm{hist}}$ converts the ordered SID interaction history into contextual representations by aggregating hierarchical SID embeddings and incorporating temporal information:
\begin{equation}
\begin{aligned}
    \mathcal{Z}^{\mathrm{h}}
    &=\mathrm{Enc}_{\mathrm{hist}}
    \left(
    [\sum_{d=1}^{D}\mathcal{E}[s_t^d]+p^{\mathrm{h}}_t]_{t=1}^{T}
    \right),
\end{aligned}
\end{equation}
where $\mathcal{E}$ denotes the shared SID embedding table and $p^{\mathrm{h}}_t$ denotes the temporal embedding at position $t$.

\subsubsection{List-Wise Preference Planner}

Rather than directly decoding item SIDs, OGR models cross-item dependencies in a coarse-grained preference space through the preference planner.
During training, the planner adopts teacher forcing. 
For each target item at position \(m\), we aggregate its \(D\)-level SID embeddings and construct the right-shifted planner input as: 
\begin{equation}
g_m=\sum_{d=1}^{D}\mathcal{E}[s_m^d],
\qquad
\mathcal{Q}^{\mathrm{tr}}
=[\mathrm{BOS},g_1,\ldots,g_{K-1}],
\label{eq:planner_input}
\end{equation}
where \(\mathcal{E}\) denotes the shared SID embedding table. A causally masked Transformer decoder then produces one planned preference representation for each slate position:
\begin{equation}
\mathcal{P}
=
\operatorname{Dec}_{\mathrm{plan}}
(\mathcal{Q}^{\mathrm{tr}},\mathcal{Z}^{\mathrm{h}})
=
[p_1,\ldots,p_K].
\label{eq:planner_output}
\end{equation}

Causal self-attention captures dependencies among preceding slate positions, while cross-attention retrieves signals relevant to the user's preferences from \(\mathcal Z^{\mathrm h}\).
During inference, the planner starts from \(\mathrm{BOS}\) and autoregressively feeds the previously generated representation \(p_{m-1}\) to produce \(p_m\).
Once \(p_m\) becomes available, the decoder starts generating the SID chain at position \(m\) while the planner proceeds to position \(m+1\), forming a pipeline between preference planning and SID decoding.




\subsubsection{Position-Wise SID Decoder}
For the \(m\)-th position, SID tokens are generated autoregressively conditioned on the planned preference embedding \(p_m\), preceding SID tokens \(s_m^{<d}\), and historical representations \(\mathcal{Z}^{\mathrm{h}}\):
\begin{equation}
\mathbf{s}_m
=
\mathrm{Dec}_{\mathrm{sid}}
\left(
p_m,\mathcal{E}[s_m^{<d}],\mathcal{Z}^{\mathrm{h}}
\right).
\label{eq:sid-decoder}
\end{equation}

Since cross-position dependencies are captured by the listwise preference planner, the SID chain at each position can be decoded independently once its planned representation \(p_m\) becomes available.
This allows current-position SID decoding to overlap with subsequent preference planning, yielding pipelined execution while preserving intra-item SID dependencies.

\subsubsection{Feedback-Aware Supervision}

We construct two complementary supervision sequences over the same target slate: 
the exposure sequence $\mathcal{Y}^{\mathrm{eps}}_u$ maintains the original display order, whereas the feedback sequence $\mathcal{Y}^{\mathrm{fb}}_u$ rearranges items according to their observed item-level feedback.
For each supervision type $o\in\{\mathrm{eps},\mathrm{fb}\}$, the SID generation loss is defined as:
\begin{equation}
    \mathcal{L}^{o}_{\mathrm{sid}}
    =-\frac{1}{|\mathcal U|KD} \sum_{u}\sum_{m=1}^{K}\sum_{d=1}^{D}
    \log \pi^{o,d}_{u,m}[s^{o,d}_{u,m}],
    \label{eq:sid-loss}
\end{equation}
where \(s^{o,d}_{u,m}\) denotes the target SID token at depth \(d\), \(\pi^{o,d}_{u,m}[s^{o,d}_{u,m}]\) denotes its predicted probability, and \(|\mathcal U|\) denotes the number of training users.

Thus, the supervised training objective is:
\begin{equation}
    \mathcal{L}_{\mathrm{sup}}
    =\mathcal{L}^{\mathrm{eps}}_{\mathrm{sid}}
    +\alpha\mathcal{L}^{\mathrm{fb}}_{\mathrm{sid}},
    \label{eq:supervised}
\end{equation}
where $\alpha$ controls the strength of feedback-aware supervision.

Under this supervision scheme, the exposure branch preserves fidelity to logged slate distributions, while the feedback branch injects user preference signals into the generation process.

\begin{table*}[t]
\centering
\setlength{\tabcolsep}{4.5pt}
\begin{tabular}{l|ccccc|ccccc}
\toprule
& \multicolumn{5}{c|}{\textbf{Industrial}}
& \multicolumn{5}{c}{\textbf{KuaiRec}} \\
\cmidrule(lr){2-6}\cmidrule(lr){7-11}
\textbf{Methods}
& \multicolumn{2}{c}{Impressions}
& \multicolumn{2}{c}{Effective Views}
& \raisebox{-1.2ex}[0pt][0pt]{NDCG@5}
& \multicolumn{2}{c}{Impressions}
& \multicolumn{2}{c}{Effective Views}
& \raisebox{-1.2ex}[0pt][0pt]{NDCG@5} \\
\cmidrule(lr){2-3}\cmidrule(lr){4-5}
\cmidrule(lr){7-8}\cmidrule(lr){9-10}
& hit@5 & recall@5 & hit@5 & recall@5 & 
& hit@5 & recall@5 & hit@5 & recall@5 & \\
\midrule
BERT4Rec
& 0.1137 & 0.0196 & 0.0689 & 0.0152 & 0.0214
& 0.3538 & 0.0908 & 0.2393 & 0.0983 & 0.0870 \\
SASRec
& 0.1206 & 0.0220 & 0.0702 & 0.0181 & 0.0223
& 0.3760 & 0.0952 & 0.2348
& 0.1013 & 0.0878 \\
Caser
& 0.1094 & 0.0164 & 0.0605 & 0.0134 & 0.0201
& 0.3329 & 0.0813 & 0.1966 & 0.0802 & 0.0765 \\
\midrule
PRM
& 0.1119 & 0.0235 & 0.0819 & 0.0201 & 0.0413
& 0.3705 & 0.0973 & 0.2470 & 0.1102 & \underline{0.1024} \\
Seq2Slate
& 0.1212 & 0.0251 & 0.0872 & 0.0220 & \underline{0.0427}
& 0.3760 & 0.0982 & 0.2500 & 0.1094 & 0.1017 \\
\midrule
TIGER
& 0.0984 & 0.0125 & 0.0517 & 0.0109 & 0.0181
& 0.3036 & 0.0779 & 0.2012 & 0.0867 & 0.0819 \\
OneRec
& \underline{0.1642} & \underline{0.0413} & \underline{0.0953} & \underline{0.0321} & 0.0394
& \underline{0.4448} & \underline{0.1105} & \underline{0.2782} & \underline{0.1117} & 0.0949 \\
\midrule
\textbf{OGR (Ours)}
& \textbf{0.2155} & \textbf{0.0511} & \textbf{0.1227}
& \textbf{0.0416} & \textbf{0.0633}
& \textbf{0.5468} & \textbf{0.1393} & \textbf{0.2991}
& \textbf{0.1475} & \textbf{0.1303} \\
\bottomrule
\end{tabular}
\caption{Offline performance of OGR and baselines on industrial and public datasets.}
\vspace{-2mm}
\label{tab:1}
\end{table*}

\subsection{Slate-Level Preference Alignment (SPA)}

Although feedback-aware supervised learning provides \textsc{OGR} with both exposure-order and feedback-informed supervision, it remains a likelihood-based imitation of fixed targets and lacks the ability to optimize relative preferences among multiple generated slates.
Inspired by preference alignment methods in large language models~\cite{DPO,InstructGPT}, we propose a preference alignment post-training stage to further align generated slates with user preferences.

\subsubsection{Candidate Rollouts}

Let $\theta_0$ denote the frozen supervised model as the reference policy.
For each user $u$, we perform beam search at each slate position using its corresponding planned preference embedding and combine the generated candidates into a non-duplicate slate-level action set $\mathcal{A}_u$.
Each action $a\in\mathcal{A}_u$ is defined as:
\[
a=(\mathbf{s}_{u,1},\ldots,\mathbf{s}_{u,K}),
\]
where $\mathbf{s}_{u,k}$ denotes the SID sequence at position $k$ and the reference log-likelihood of a candidate slate is computed as:
\begin{equation}
\ell_0(a|u)
=
\sum_{m=1}^{K}
\sum_{d=1}^{D}
\log
p_{\theta_0}
\left(
s^{d}_{u,m}
\mid
\mathcal{Z}^{\mathrm{h}}_u,
p_{u,m},
s^{<d}_{u,m}
\right),
\label{eq:old-score}
\end{equation}
where $p_{u,m}$ denotes the position-aware preference embedding generated by the List-Wise Preference Planner.

\subsubsection{Reward-Consistent Calibration}

To evaluate candidate slates from multiple perspectives, we define a primary reward \(r_{\mathrm{pri}}\) that measures slate-level user preference by aggregating feedback signals over the entire slate, such as effective views and likes, and an auxiliary reward \(r_{\mathrm{aux}}\) that provides additional guidance through slate-level quality signals, such as diversity. The details are provided in the appendix. 

For $o\in\{\mathrm{pri},\mathrm{aux}\}$, we standardize rewards within $\mathcal{A}_u$ to obtain the relative reward score $\delta^o_{u,a}$, and the final calibrated preference signal $\Delta_{u,a}$ is computed as:
\begin{equation}
\begin{aligned}
\kappa_{u,a}
&=
\begin{cases}
\displaystyle
\frac{
\min\big(
|\delta^{\mathrm{pri}}_{u,a}|,
|\delta^{\mathrm{aux}}_{u,a}|
\big)}
{
\max\big(
|\delta^{\mathrm{pri}}_{u,a}|,
|\delta^{\mathrm{aux}}_{u,a}|,
\varepsilon
\big)
},
&
\delta^{\mathrm{pri}}_{u,a} \cdot \delta^{\mathrm{aux}}_{u,a}>0,
\\[8pt]
0,
&
\mathrm{otherwise},
\end{cases}
\\
\Delta_{u,a}
&=
\delta^{\mathrm{pri}}_{u,a}
+\kappa_{u,a}\delta^{\mathrm{aux}}_{u,a}.
\end{aligned}
\label{eq:feedback-calibration}
\end{equation}

Here, $\kappa_{u,a}$ scales the auxiliary contribution according to the relative score magnitudes and suppresses it when the primary and auxiliary scores have conflicting directions, and $\varepsilon>0$ avoids division by zero.
Consequently, the calibration preserves the primary recommendation objective while incorporating reliable auxiliary signals, resulting in a more robust preference alignment process.

\subsubsection{Conservative Policy Optimization}

Given the same candidate action sets, we compute the importance ratio
\(\rho_{u,a}=\exp(\ell_{\theta}(a\mid u)-\ell_0(a\mid u))\)
and optimize the clipped policy objective~\cite{PPO}:
\begin{equation}
\begin{aligned}
\mathcal{L}_{\mathrm{pol}}
&=
-\frac{1}{N_A}
\sum_{u}\sum_{a\in\mathcal{A}_u}
\min\Bigl\{
\rho_{u,a}\Delta_{u,a},\\
&\qquad\qquad
\operatorname{clip}\!\left(\rho_{u,a},1-\epsilon,1+\epsilon\right)
\Delta_{u,a}
\Bigr\}.
\end{aligned}
\end{equation}

where \(N_A\) denotes the number of retained user-action pairs and \(\epsilon\) is the clipping radius.

To further constrain policy drift, we derive the candidate slate distributions from the reference and current policy likelihoods:
$q_0(a\mid u)=\mathrm{softmax}_{a\in\mathcal A_u}(\ell_0(a\mid u))$
and
$q_\theta(a\mid u)=\mathrm{softmax}_{a\in\mathcal A_u}(\ell_\theta(a\mid u))$
. Their divergence is then defined as:
\begin{equation}
    \mathcal{L}_{\mathrm{KL}}
    =\frac{1}{N_U}\sum_u\sum_{a\in\mathcal{A}_u}
    q_0(a\mid u)\log\frac{q_0(a\mid u)}{q_{\theta}(a\mid u)},
    \label{eq:candidate-kl}
\end{equation}
where $N_U$ denotes the number of retained users.
We additionally replay the supervised objective $\mathcal{L}_{\mathrm{sup}}$ and the final alignment objective is as follows:
\begin{equation}
    \mathcal{L}_{\mathrm{SPA}}
    =\mathcal{L}_{\mathrm{pol}}
    +\gamma\mathcal{L}_{\mathrm{KL}}
    +\eta\mathcal{L}_{\mathrm{sup}}.
    \label{eq:refinement}
\end{equation}
where $\gamma$ and $\eta$ denote the corresponding weights.

During refinement, we update only the List-Wise Preference Planner and the Position-Wise SID Decoder, while keeping remaining model components frozen. Consequently, slate-level planning and item-level decoding are jointly aligned with user feedback while avoiding excessive deviation from the supervised generation policy.

\section{Experiment}

\subsection{Experimental Settings}

\subsubsection{Datasets and Evaluation Metrics}

We conduct experiments on KuaiRec~\cite{KuaiRec} and a proprietary industrial dataset collected from Kuaishou. The industrial dataset contains interaction sequences, multimodal content, high-level attributes, exposure logs, and item-level feedback. For both datasets, interactions are chronologically ordered and evaluated using a leave-five-out protocol. We report Hit Rate, Recall, and NDCG~\cite{NDCG} on positive-feedback lists to assess the model's ability to capture users' preference, and additionally report impression-based Hit Rate and Recall on logged exposures to evaluate the model's ability to capture observed exposure patterns.

\subsubsection{Implementation Details}

All offline experiments are conducted on NVIDIA L20 GPUs. We use a maximum history length of 128 and generate slates of size 5. The SID tokenizer adopts a four-level configuration with a codebook size of 1,024 at each level. During inference, we perform beam search with a beam width of 20. Additional implementation details are provided in the appendix.

\subsection{Results}

\subsubsection{Overall Performance Comparison}

As shown in Table~\ref{tab:1}, we compare \textsc{OGR} against three categories of baselines: sequential recommenders (Caser, SASRec, and BERT4Rec)~\cite{Caser,SASRec,BERT4Rec}, listwise rerankers (PRM and Seq2Slate)~\cite{PRM,Seq2Slate}, and generative recommenders (TIGER and OneRec)~\cite{TIGER,OneRec}. All generative baselines are implemented with comparable parameter sizes to OGR (32 MB).

Discriminative sequential recommendation methods directly estimate item scores and select the highest-scoring items, leading to stable performance in Top-5 recommendation.
Listwise reranking methods explicitly model interactions among items within a candidate set and therefore provide stronger slate-level ordering ability.
Meanwhile, although OneRec achieves strong Hit Rate and Recall, the method yields lower NDCG than listwise reranking baselines, highlighting a decoding dilemma: autoregressive slate generation captures cross-item dependencies but suffers from serial latency and error accumulation, whereas probability-based Top-5 selection is efficient but not listwise-aligned under NTP loss.
In contrast, \textsc{OGR} addresses this dilemma by combining global slate planning with pipelined position-wise SID decoding, preserving cross-position dependencies while avoiding serial item-by-item SID generation and thereby achieving consistent gains across all evaluation metrics.

\subsubsection{Comparison of SID Construction Methods}

\begin{table*}[t]
\centering
{\small
\setlength{\tabcolsep}{5pt}
\begin{tabular}{l|cc|cccc|cc}
\toprule
\multicolumn{1}{c|}{\raisebox{-2ex}[0pt][0pt]{\textbf{SID Method}}}
& \multicolumn{2}{c|}{\shortstack{\textbf{Downstream}\\\textbf{Recommendation}}}
& \multicolumn{4}{c|}{\raisebox{1ex}[0pt][0pt]{\textbf{Codebook Balance}}}
& \multicolumn{2}{c}{\shortstack{\textbf{Content}\\\textbf{Preservation}}} \\
\cmidrule(lr){2-3}\cmidrule(lr){4-7}\cmidrule(lr){8-9}
& \shortstack{Eff.\\Recall@5 $\uparrow$}
& \raisebox{1ex}[0pt][0pt]{NDCG@5 $\uparrow$}
& \raisebox{1ex}[0pt][0pt]{ICR $\uparrow$}
& \raisebox{1ex}[0pt][0pt]{CUR $\uparrow$}
& \shortstack{Min.\\PPL $\uparrow$}
& \shortstack{Top-1\\Load $\downarrow$}
& \shortstack{V-measure\\Level-1 $\uparrow$}
& \raisebox{1ex}[0pt][0pt]{SC $\uparrow$} \\
\midrule
RQ-VAE (TIGER)
& 0.117 & 0.112
& 0.996 & 0.752 & 7.8 & 0.047
& 0.305 & 0.622 \\
RQ-KMeans (OneRec)
& 0.121 & \underline{0.116}
& 0.995 & \textbf{1.000} & 468.1 & 0.028
& \textbf{0.431} & 0.737 \\
R3-VAE
& 0.120 & 0.114
& 0.984 & 0.683 & 12.4 & 0.041
& 0.269 & 0.619 \\
GNPR-SID
& \underline{0.129} & 0.115
& \textbf{1.000} & \underline{0.955}
& \textbf{836.8} & \textbf{0.004}
& 0.412 & \underline{0.741} \\
LC-Rec
& 0.116 & 0.111
& \textbf{1.000} & 0.764 & 52.9 & 0.011
& 0.381 & 0.660 \\
LETTER
& 0.114 & 0.109
& \underline{0.999} & 0.335 & 39.8 & 0.030
& 0.241 & 0.608 \\
\midrule
\textbf{TUSID (Ours)}
& \textbf{0.148} & \textbf{0.130}
& \textbf{1.000} & \textbf{1.000}
& \underline{681.7} & \underline{0.005}
& \underline{0.424} & \textbf{0.747} \\
\bottomrule
\end{tabular}
}
\caption{Comparison of SID methods on KuaiRec across three evaluation dimensions.}
\vspace{-2.5mm}
\label{tab:2}
\end{table*}

\begin{table*}[t]
\centering
{\small
\setlength{\tabcolsep}{2.8pt}
\begin{tabular}{
>{\raggedright\arraybackslash}m{1.05in}|
>{\centering\arraybackslash}m{0.62in}
>{\centering\arraybackslash}m{0.88in}
>{\centering\arraybackslash}m{0.90in}
>{\centering\arraybackslash}m{0.80in}
>{\centering\arraybackslash}m{0.40in}|
>{\centering\arraybackslash}m{0.85in}
>{\centering\arraybackslash}m{0.68in}}
\toprule
\textbf{Variant}
& \shortstack{\textbf{High-level}\\\textbf{Attr.}}
& $e^{\mathrm{sem}}$ \textbf{Fusion}
& $e^{\mathrm{col}}$
& $e^{\mathrm{col}}$ \textbf{Fusion}
& \textbf{CAW}
& \textbf{Eff. Recall@5}
& \textbf{NDCG@5} \\
\midrule

\shortstack[l]{Base SID\\(RQ-KMeans)}
& $\times$ 
& N/A 
& N/A 
& N/A 
& N/A
& 0.1214 
& 0.1162 \\

\midrule

\multirow[c]{2}{*}{\raisebox{-0.8\height}{\shortstack[l]{+ Semantic\\Fusion}}}
& $\checkmark$ 
& Add 
& N/A 
& N/A 
& N/A
& 0.1138 
& 0.1096 \\

& $\checkmark$ 
& \shortstack{Cross-Attn\\\& Gate}
& N/A 
& N/A 
& N/A
& 0.1295 
& 0.1198 \\

\midrule

\multirow[c]{3}{*}{\raisebox{-1.3\height}{\shortstack[l]{+ Collaborative\\Injection}}}
& $\checkmark$ 
& \shortstack{Cross-Attn\\\& Gate}
& CCE 
& Add 
& $\times$
& 0.1387 
& 0.1248 \\

& $\checkmark$ 
& \shortstack{Cross-Attn\\\& Gate}
& InfoNCE Loss 
& N/A 
& $\times$
& 0.1404 
& 0.1275 \\

& $\checkmark$ 
& \shortstack{Cross-Attn\\\& Gate}
& CCE 
& Concat 
& $\times$
& \underline{0.1437} 
& \underline{0.1298} \\

\midrule

\textbf{TUSID (Full)}
& $\checkmark$ 
& \shortstack{Cross-Attn\\\& Gate}
& CCE 
& Concat 
& $\checkmark$
& \textbf{0.1475} 
& \textbf{0.1303} \\

\bottomrule
\end{tabular}
}
\caption{Ablation study of the proposed components in TUSID on KuaiRec. Here, $e^{\mathrm{sem}}$ denotes the attribute-enriched semantic embedding, and $e^{\mathrm{col}}$ denotes the collaborative embedding. InfoNCE Loss~\cite{InfoNCE} denotes the contrastive variant that aligns semantic embeddings using item co-occurrence-based positive and negative pairs directly.}
\vspace{-3.5mm}
\label{tab:3}
\end{table*}

Table~\ref{tab:2} compares SID methods~\cite{TIGER,OneRec,R3VAE,GNPRSID,LCRec,LETTER} under the same \textsc{OGR} backbone and codebook size.

TUSID consistently outperforms all public baselines on both downstream recommendation metrics, which indicates that the resulting SIDs more effectively capture user preferences and collaborative structures.
ICR and CUR measure SID uniqueness and codebook coverage~\cite{OneSearch}, while Min. PPL and Top-1 Load evaluate code assignment balance by reflecting code utilization and concentration in dominant codes~\cite{AdaSID}.
The results show that TUSID produces collision-free SIDs while maintaining balanced code assignments across the hierarchical levels. 
For semantic preservation, TUSID maintains comparable V-measure~\cite{VMeasure} and SC~\cite{R3VAE}, which evaluate category consistency and semantic cohesion, respectively.
This result suggests that effective SIDs require a balance of semantic preservation, collaborative structure, and codebook quality rather than optimizing a single metric.

\subsubsection{Ablation Study of SID Construction}

Table~\ref{tab:3} further investigates the contribution of each component in TUSID.
Compared with cross-attention and gated fusion, direct addition of high-level attributes yields lower performance, indicating that naive fusion of heterogeneous semantic sources can disrupt the original semantic structure.
For collaborative injection, CCE with concatenation outperforms both additive fusion and InfoNCE-based collaborative supervision, showing that preserving complementary semantic and collaborative structures is more effective than collapsing them into a shared space. 
CAW further improves performance, demonstrating that confidence-based scaling of collaborative information helps prevent sparse or accidental interactions from being amplified into reliable preference signals.

\begin{table}[t]
\centering
{\small
\renewcommand{\arraystretch}{1.08}
\setlength{\tabcolsep}{3pt}
\begin{tabular}{
>{\raggedright\arraybackslash}m{1.86in}|
>{\centering\arraybackslash}m{0.62in}
>{\centering\arraybackslash}m{0.58in}}
\toprule
\textbf{Variant}
& \shortstack{\textbf{Eff.}\\\textbf{Recall@5}}
& \textbf{NDCG@5} \\
\midrule
\textbf{GL2P+SPA (Full)}
& \textbf{0.0416}
& \textbf{0.0633} \\
\midrule
w/o List-Wise Planner
& 0.0201
& 0.0359 \\
w/o SPA
& 0.0344
& 0.0419 \\
\midrule
w/o Primary Reward
& 0.0332
& 0.0467 \\
w/o Auxiliary Reward
& 0.0403
& 0.0554 \\
w/o Conservative Regularization
& 0.0227
& 0.0312 \\
\bottomrule
\end{tabular}
}
\caption{Ablation study of the proposed GL2P and SPA on Industrial. All variants use the same TUSID tokenizer.}
\vspace{-5mm}
\label{tab:generation_ablation}
\end{table}

\subsubsection{Ablation Study of Slate Generation and Alignment}



Table~\ref{tab:generation_ablation} evaluates the major components of GL2P and SPA.
Removing the List-Wise Planner substantially degrades both metrics despite reallocating its parameter budget to the Encoder/Decoder, demonstrating that global preference planning is essential for coordinating item selection and slate ordering.
Removing SPA also hurts performance, indicating that supervised generation alone is insufficient to optimize user-preferred slates.

Within SPA, removing the primary reward causes a larger degradation than removing the auxiliary reward, confirming that user feedback provides the dominant alignment signal while slate-level quality signals offer complementary guidance. 
The full model achieves the best performance by combining both signals with confidence-aware calibration.
Finally, removing conservative regularization results in a significant decline, validating the importance of constraining policy deviation during preference alignment.

Table~\ref{tab:supervision_ablation} further examines the supervision objectives in GL2P. Removing either $\mathcal{L}^{\mathrm{eps}}_{\mathrm{sid}}  $ or $ \mathcal{L}^{\mathrm{fb}}_{\mathrm{sid}} $ reduces both metrics, indicating that exposure-order supervision provides structural guidance for slate generation, while feedback-aware supervision refines generation according to user responses. Only their joint supervision enables OGR to learn stable and effective slate-level sequential dependencies.



\begin{table}[t]
\centering
{\small
\renewcommand{\arraystretch}{1.08}
\setlength{\tabcolsep}{3pt}
\begin{tabular}{
>{\raggedright\arraybackslash}m{1.52in}|
>{\centering\arraybackslash}m{0.82in}
>{\centering\arraybackslash}m{0.62in}}
\toprule
\textbf{Variant}
& \shortstack{\textbf{Effective-View}\\\textbf{Recall@5}}
& \textbf{NDCG@5} \\
\midrule
$ \mathcal{L}_{\mathrm{sup}} = \mathcal{L}^{\mathrm{eps}}_{\mathrm{sid}} + \alpha\mathcal{L}^{\mathrm{fb}}_{\mathrm{sid}} $
& \textbf{0.1475}
& \textbf{0.1303} \\
w/o $\mathcal{L}^{\mathrm{eps}}_{\mathrm{sid}}  $
& 0.1382
& 0.1201 \\
w/o $ \mathcal{L}^{\mathrm{fb}}_{\mathrm{sid}} $
& 0.1230
& 0.1139 \\
\bottomrule
\end{tabular}
}
\caption{Ablation study of the supervision objectives in GL2P on
KuaiRec. All other generation and alignment components are fixed.}
\label{tab:supervision_ablation}
\end{table}


\subsubsection{Efficiency Analysis}

\begin{figure}[t]
  \centering
  \IfFileExists{Figures/five_decode_modes_efficiency_prediction.pdf}{%
    \includegraphics[width=\linewidth]{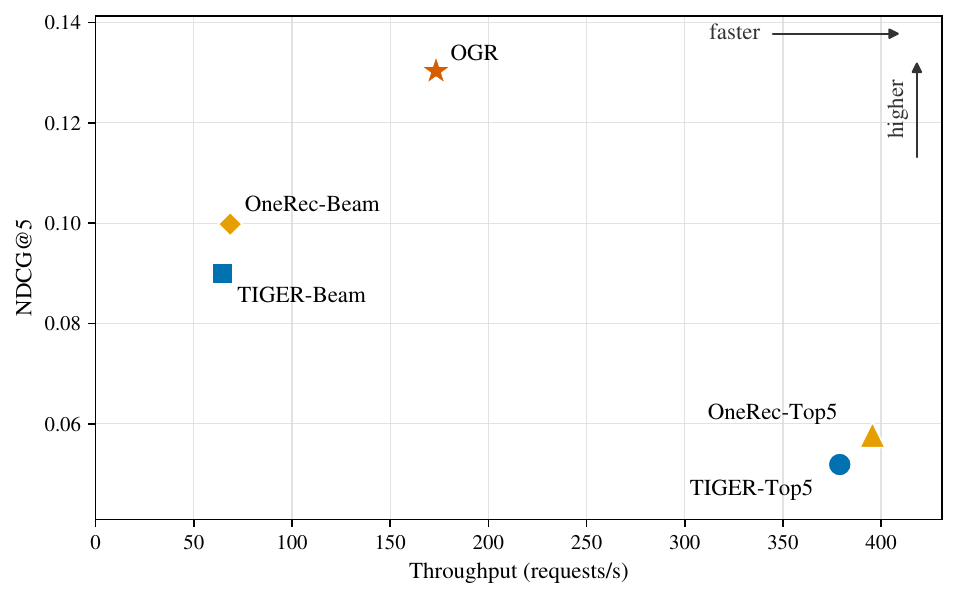}%
  }{%
    \fbox{\parbox[c][1.8in][c]{0.94\linewidth}{\centering
      Figure file unavailable.\\[3pt]
      Add the referenced PDF under the \texttt{Figures} directory.}}%
  }
  \caption{Efficiency comparison under Beam-5 and Top-5 decoding modes for TIGER and OneRec, together with OGR.}
  \label{fig:decoding-efficiency}
\end{figure}

For a fair and comprehensive efficiency comparison, we consider two decoding modes for the generative baselines in Fig.~\ref{fig:decoding-efficiency}.
Beam-5 autoregressively generates a $K$-item slate by serially decoding the $D$-level SID of each slate position.
Because all $K\times D$ SID tokens are generated
one after another, completing the slate requires $KD$ consecutive
autoregressive decoding steps, resulting in the sequential decoding depth $\mathrm{Depth}_{\mathrm{Beam\text{-}5}}=KD$.
In contrast, Top-5 performs a single next-item beam search and directly returns the \(K\) highest-scoring item candidates.
The SID chains of all \(K\) items are decoded in parallel within the same beam search, so the sequential decoding depth depends only on the \(D\) intra-item decoding steps, giving \(\mathrm{Depth}_{\mathrm{Top\text{-}5}}=D\).

In \textsc{OGR}, the efficiency improvement arises from the reduction in sequential dependencies.
Once the planned representation $p_m$ is available, $\mathrm{Dec}_{\mathrm{sid}}$ can generate its SID chain while $\mathrm{Dec}_{\mathrm{plan}}$ continues producing subsequent preference representations.
For position $m$, the $d$-th SID token $s_m^d$ becomes available after $m$ planning steps and $d$ intra-item SID-decoding steps, so
its sequential dependency depth is $m+d$.
The complete slate is available when the final token $s_K^D$ is generated, giving $\mathrm{Depth}_{\mathrm{OGR}}=K+D$.

Accordingly, \textsc{OGR} reduces the sequential dependency depth
from $\mathcal{O}(KD)$ to $\mathcal{O}(K+D)$ by overlapping each position's SID decoding with preference planning for subsequent positions, while preserving cross-position dependencies through $\mathrm{Dec}_{\mathrm{plan}}$.
Under our setting of $K=5$ and $D=4$, the idealized critical-path reduction factor is $KD/(K+D)=20/9\approx2.22$.
This shortened sequential path provides a structural explanation for the observed acceleration.
As Fig.~\ref{fig:decoding-efficiency} shows, \textsc{OGR} achieves measured throughput improvements of $2.43\times$ and $2.49\times$ over TIGER-Beam and OneRec-Beam, respectively, while obtaining the highest NDCG@5.
By contrast, the Top-5 variants require only one next-item beam search but achieve substantially lower recommendation quality because they return alternative next-item candidates without explicitly modeling dependencies and ordering preferences across slate positions.
Consequently, these results suggest that \textsc{OGR} provides an effective
quality--efficiency trade-off for generative slate recommendation.

\subsubsection{Online A/B Testing}
We conduct online A/B testing on Kuaishou with 3\% of the production traffic. Compared with the existing system, \textsc{OGR} improves Effective Views, Comments, Likes, and Forwards by 1.120\%, 2.954\%, 0.505\%, and 1.255\%, respectively, demonstrating its effectiveness and practical value in real-world large-scale recommendation scenarios.

\section{Conclusion}

We propose OGR, an end-to-end generative framework for slate recommendation that realizes ``\textbf{Once Generated, Ranked}.'' 
OGR integrates TUSID for constructing recommendation-aware hierarchical SIDs and a list-wise generation architecture with preference alignment for directly producing ordered slates. 
Extensive offline experiments and online A/B testing demonstrate its effectiveness and practical impact in real-world recommendation scenarios.

\bibliography{aaai2027}

\clearpage
\appendix

\begin{center}
{\LARGE\bfseries Technical Appendix\par}
\end{center}
\vspace{1em}

\section{Appendix 0: Appendix Overview}

This section provides an overview of the supplementary material
and its correspondence to the main paper.

\begin{center}
{\footnotesize
\renewcommand{\arraystretch}{1.03}
\setlength{\tabcolsep}{1mm}
\begin{tabular}{@{}>{\raggedright\arraybackslash}p{0.17\columnwidth}
                    >{\raggedright\arraybackslash}p{0.47\columnwidth}
                    >{\raggedright\arraybackslash}p{0.28\columnwidth}@{}}
\toprule
Appendix & Contents & Main-paper subsection \\
\midrule
Appendix~1 &
Additional Related Work on SID Construction &
Related Works \\
Appendix~2 &
B-Tags and C-Tags &
Semantic Fusion \\
Appendix~3 &
RQ-KMeans for SID Construction &
Collaborative Injection \\
Appendix~4 &
Details of Collaborative Injection &
Collaborative Injection \\
Appendix~5 &
Details of Reward-Consistent Calibration &
Reward-Consistent Calibration \\
Appendix~6 &
Hyperparameter Configuration &
Implementation Details \\
Appendix~7 &
Symbol Glossary &
Main paper \\
\bottomrule
\end{tabular}\par}
\captionof{table}{Overview of the supplementary material
and its correspondence to the main paper.}
\label{tab:item-directory}
\end{center}

\section{Appendix 1: Additional Related Work on SID Construction}

The related work on SID construction methods is presented below:

Traditional recommender systems have long represented items using atomic IDs,
a paradigm widely adopted in collaborative filtering and factorization methods
(ItemKNN, MF, BPR-MF, and FISM)
\citep{sarwar2001itemknn,koren2009mf,rendle2009bpr,kabbur2013fism},
deep sequential models (GRU4Rec, Caser, NARM, and NextItNet)
\citep{hidasi2016gru4rec,tang2018caser,li2017narm,yuan2019nextitnet}, and
Transformer-based recommenders (SASRec, BERT4Rec, and S3-Rec)
\citep{kang2018sasrec,sun2019bert4rec,zhou2020s3rec}.
However, atomic IDs are arbitrary and unstructured symbols that encode no
shared item structure, hindering the effective modeling of sparsely observed,
long-tail, and newly introduced items.
To address this limitation, TIGER~\citep{rajput2023tiger} uses RQ-VAE to
quantize content representations into hierarchical SIDs, while QARM and
OneRec~\citep{luo2025qarm,deng2025onerec,zhou2025onerectr} adopt RQ-KMeans to
improve codebook utilization and assignment balance.
These studies establish the foundation for SIDs in generative recommendation,
yet the effective integration of content semantics and local collaborative
structures remains underexplored.

Existing semantic-collaborative SID construction methods broadly follow two
directions.
The first directly imposes collaborative constraints during quantization, as
in LETTER, ETEGRec, HiGR, QuaSID, and
AdaSID~\citep{wang2024letter,liu2025etegrec,pang2026higr,hu2026quasid,
pan2026adasid}.
However, jointly optimizing collaborative alignment with reconstruction and
codebook objectives may introduce competing gradients, increasing the risk of
training instability and codebook collapse.
The second aligns semantic and behavioral representations in the continuous
space before quantization, as in QARM, FORGE, OneSearch, SaviorRec, and
GR4AD~\citep{luo2025qarm,fu2026forge,chen2026onesearch,yao2025saviorrec,
xue2026gr4ad}.
These approaches decouple representation learning from discrete encoding but
typically rely on sampled item pairs and contrastive learning, making it
difficult to fully capture local co-occurrence structures of varying strengths
and limiting their suitability for generative slate recommendation.

\section{Appendix 2: B-Tags and C-Tags}

\begin{table*}[t]
\centering
\small
\setlength{\tabcolsep}{2mm}
\begin{tabular}{lll}
\toprule
Type & Dimension & Examples \\
\midrule
\multirow{8}{*}{B-tags (supply side)}
& Positioning & knowledge sharing; entertainment; daily life; product promotion \\
& Persona & expert; friend-like creator; comedian; reviewer \\
& Monetization & product/brand promotion; traffic acquisition; course conversion \\
& Format & narration; plot; review; tutorial; interview; Vlog \\
& Style & professional; humorous; authentic; premium; fast-paced \\
& Emotion & joyful; warm; healing; tense; encouraging \\
& Depth & entertainment; introduction; practical explanation; in-depth analysis \\
& Operation & trend leverage; suspense hook; interaction prompt; serialization \\
\midrule
\multirow{4}{*}{C-tags (demand side)}
& Pain Point & makeup difficulty; low efficiency; weight loss; product selection \\
& Benefit & knowledge; problem solving; relaxation; discounts; efficiency \\
& Scenario & commute; bedtime; home; office; travel; purchase decision \\
& Audience & students; professionals; mothers; novices; digital enthusiasts \\
\bottomrule
\end{tabular}
\caption{Taxonomy of B-tags and C-tags. Each dimension may contain multiple values.}
\label{tab:bc-tags}
\end{table*}

In addition to Brand and Category, we use two multi-label
high-level attributes for each item, as shown in
Table~\ref{tab:bc-tags}. B-tags characterize the
\emph{supply side} of content, including how it is positioned,
presented, and operated; C-tags characterize the \emph{demand side},
including the user need, value, scenario, and target audience.
The two tag records are separately converted into text and encoded by
the shared text encoder to obtain $e_i^{\mathrm{Bt}}$ and
$e_i^{\mathrm{Ct}}$, which are fused with the multimodal item
representation as described in the main text.

\section{Appendix 3: RQ-KMeans for SID Construction}

Given the unified item representation \(e_i^{\mathrm{uni}}\), we apply
residual-quantized K-means (RQ-KMeans) to construct the hierarchical SID
\(\mathbf{s}_i\). RQ-KMeans recursively quantizes the residual representation,
so that early codewords typically capture larger-scale variation, whereas later
codewords encode the remaining residual detail.

Let \(\zeta_i^{(1)}=e_i^{\mathrm{uni}}\) be the initial residual of item \(i\).
Here, \(i\in\{1,\ldots,N_{\mathrm{item}}\}\), \(D\) is the SID depth used in
the main paper, \(n_d^{\mathrm{cb}}\) is the codebook size at depth \(d\), and
\[
\Xi^{(d)}
=
\{\xi_n^{(d)}\}_{n=1}^{n_d^{\mathrm{cb}}}
\]
denotes the depth-\(d\) codebook. RQ-KMeans assigns the residual to its nearest
centroid:
\begin{equation}
s_i^d
=
\arg\min_{n\in\{1,\ldots,n_d^{\mathrm{cb}}\}}
\left\lVert
\zeta_i^{(d)}-\xi_n^{(d)}
\right\rVert_2^2.
\label{eq:appendix-rq-assignment}
\end{equation}
The residual is then updated as
\begin{equation}
\zeta_i^{(d+1)}
=
\zeta_i^{(d)}-\xi_{s_i^d}^{(d)}.
\label{eq:appendix-rq-residual}
\end{equation}
After \(D\) quantization levels, the SID and reconstructed representation are
\begin{equation}
\mathbf{s}_i=(s_i^1,\ldots,s_i^D),
\qquad
\widehat e_i^{\mathrm{rq}}
=
\sum_{d=1}^{D}\xi_{s_i^d}^{(d)}.
\label{eq:appendix-rq-reconstruction}
\end{equation}

During validation, testing, and online inference, the codebooks are frozen and
each item is encoded by recursively selecting the nearest centroid at every
depth. This ensures a consistent SID vocabulary across training and
evaluation.

\section{Appendix 4: Details of Collaborative Injection}

\subsection{Preliminaries of CountSketch}

CountSketch was originally introduced for frequency estimation in data
streams \citep{charikar2002countsketch}. Given a frequency vector
$x\in\mathbb{R}^{n_{\mathrm{vocab}}}$, it maintains a compressed table
$\mathbf G_{\mathrm{cs}}\in
\mathbb{R}^{n_{\mathrm{row}}\times n_{\mathrm{bucket}}}$.
For row $v$, a bucket hash
$\mathfrak h_v:[n_{\mathrm{vocab}}]\rightarrow[n_{\mathrm{bucket}}]$
and a sign hash
$\varsigma_v:[n_{\mathrm{vocab}}]\rightarrow\{-1,+1\}$
map a weighted update $(j,\upsilon)$ to
\begin{equation}
\mathbf G_{\mathrm{cs},v,\mathfrak h_v(j)}
\leftarrow
\mathbf G_{\mathrm{cs},v,\mathfrak h_v(j)}
+
\varsigma_v(j)\upsilon,
\qquad v=1,\ldots,n_{\mathrm{row}}.
\label{eq:countsketch-update}
\end{equation}
The corresponding row estimate is
\begin{equation}
\widehat{x}_j^{(v)}
=
\varsigma_v(j)\mathbf G_{\mathrm{cs},v,\mathfrak h_v(j)}.
\label{eq:countsketch-row-query}
\end{equation}
For independent sign hashes, cross terms caused by unrelated coordinates have
zero expectation; additional buckets reduce collisions, while the median
across independent rows makes classical point queries robust to an occasional
large collision.

CCE uses only the two feature-mapping operations needed for collaborative
representation learning: fixed-dimensional bucket hashing and random sign
hashing. It shares one bucket hash and one sign hash across all center items
and constructs an \(n_{\mathrm{bucket}}\)-dimensional sketch for each item.
Unlike classical CountSketch, CCE does not recover individual coordinates or
aggregate multiple point-query estimates. It directly applies a signed-log
transform, normalization, and random projection to the shared-hash sketch.
This design requires one bucket update per observed context pair and reduces
the item--context storage to a fixed-dimensional representation per item.

\begin{algorithm}[!tb]
\small
\caption{Collaborative embedding construction and confidence-aware fusion}
\label{alg:cce}
\textbf{Input}: Training prefixes $\{\mathcal R_u^+\}$; semantic embeddings
$\{e_i^{\mathrm{sem}}\}$\\
\textbf{Parameter}: $\omega,n_{\mathrm{bucket}},h,\sigma,\Lambda_{\mathrm{col}},\tau,
\alpha_{\mathrm{col}},\vartheta_{\mathrm n}$\\
\textbf{Output}: $\{e_i^{\mathrm{col}}\}$,
$\{e_i^{\mathrm{uni}}\}$, and $\{U_i\}$
\begin{algorithmic}[1]
\STATE Initialize $\mathcal C_i\leftarrow0_{n_{\mathrm{bucket}}}$ and $U_i\leftarrow0$
for all items $i$.
\FOR{each user $u$}
    \STATE $\mathcal I_u^{\mathrm{ctx}}\leftarrow\emptyset$
    \FOR{$p=1$ to $|\mathcal R_u^+|$}
        \STATE $i\leftarrow \mathcal R_u^+[p]$
        \FOR{$q=\max(1,p-\omega)$ to $\min(|\mathcal R_u^+|,p+\omega)$}
            \IF{$q\ne p$}
                \STATE $j\leftarrow \mathcal R_u^+[q]$,
                $\upsilon\leftarrow1/|p-q|$
                \STATE $\mathcal C_i[h(j)]\leftarrow
                \mathcal C_i[h(j)]+\sigma(j)\upsilon$;
                $\mathcal I_u^{\mathrm{ctx}}\leftarrow \mathcal I_u^{\mathrm{ctx}}\cup\{i\}$
            \ENDIF
        \ENDFOR
    \ENDFOR
    \STATE $U_i\leftarrow U_i+1$ for each $i\in \mathcal I_u^{\mathrm{ctx}}$
\ENDFOR
\FOR{each item $i$}
    \STATE $z_i[\nu]\leftarrow
    \operatorname{sign}(\mathcal C_i[\nu])
    \log(1+|\mathcal C_i[\nu]|)$ for all $\nu$
    \STATE $\bar z_i\leftarrow
    z_i/\max(\lVert z_i\rVert_2,\vartheta_{\mathrm n})$
    \STATE $\widetilde e_i^{\mathrm{col}}\leftarrow \Lambda_{\mathrm{col}}\bar z_i$;
    $e_i^{\mathrm{col}}\leftarrow
    \widetilde e_i^{\mathrm{col}}/
    \max(\lVert\widetilde e_i^{\mathrm{col}}\rVert_2,
    \vartheta_{\mathrm n})$
    \STATE $\bar e_i^{\mathrm{sem}}\leftarrow
    e_i^{\mathrm{sem}}/
    \max(\lVert e_i^{\mathrm{sem}}\rVert_2,
    \vartheta_{\mathrm n})$
    \STATE $\gamma_i\leftarrow
    \log(1+U_i)/[\log(1+U_i)+\tau]$;
    $\alpha_i\leftarrow\alpha_{\mathrm{col}}\gamma_i$
    \STATE $e_i^{\mathrm{uni}}\leftarrow
    [\sqrt{1-\alpha_i}\bar e_i^{\mathrm{sem}};
    \sqrt{\alpha_i}e_i^{\mathrm{col}}]$
\ENDFOR
\STATE \textbf{return}
$\{e_i^{\mathrm{col}}\}$,
$\{e_i^{\mathrm{uni}}\}$, and $\{U_i\}$
\end{algorithmic}
\end{algorithm}

\subsection{CountSketch-Based Collaborative Embedding}

Building on the bucket-hashing and sign-hashing operations summarized above,
CCE uses signed hashing as a
fixed-dimensional feature map for item-level collaborative structures.

Let
$\mathcal R_u^+=(\jmath_{u,1},\ldots,\jmath_{u,L_u^+})$
denote the high-quality positive-feedback training prefix of user $u$.
Validation and test interactions are excluded from collaborative feature
construction. For each item $i$, we define its uncompressed local-context
vector $x_i\in\mathbb{R}^{n_{\mathrm{vocab}}}$ as
\begin{equation}
x_i[j]
=
\sum_u
\sum_{\substack{
p,q:\;\jmath_{u,p}=i,\;\jmath_{u,q}=j\\
0<|p-q|\leq\omega
}}
\frac{1}{|p-q|},
\label{eq:context-vector}
\end{equation}
where $n_{\mathrm{vocab}}$ is the item vocabulary size and $\omega$ is the context-window
radius. The value $x_i[j]$ measures the distance-weighted local co-occurrence
between center item $i$ and contextual item $j$, with nearby interactions
receiving larger weights. When the same item appears at different sequence
positions, $j=i$ is permitted; such repeated-item co-occurrences are retained,
while only the identical position $q=p$ is excluded.

Explicitly storing all $\{x_i\}$ requires an item--context matrix whose size
grows with the item vocabulary. CCE instead employs one bucket hash
\begin{equation}
h:[n_{\mathrm{vocab}}]\rightarrow[n_{\mathrm{bucket}}]
\end{equation}
and one sign hash
\begin{equation}
\sigma:[n_{\mathrm{vocab}}]\rightarrow\{-1,+1\},
\end{equation}
which are shared by all center items. The compressed collaborative sketch of
item $i$ is defined as
\begin{equation}
\mathcal{C}_i[\nu]
=
\sum_{j:h(j)=\nu}
\sigma(j)x_i[j],
\qquad \nu=1,\ldots,n_{\mathrm{bucket}}.
\label{eq:cce-sketch}
\end{equation}
Equivalently, when contextual item $j=\jmath_{u,q}$ occurs within the local
window of center item $i=\jmath_{u,p}$, the corresponding sketch is updated by
\begin{equation}
\mathcal{C}_i[h(j)]
\leftarrow
\mathcal{C}_i[h(j)]
+
\frac{\sigma(j)}{|p-q|}.
\label{eq:cce-stream-update}
\end{equation}

Sharing $h$ and $\sigma$ across all center items makes their compressed
contextual distributions directly comparable. In particular, the raw
sketches satisfy
\begin{equation}
\mathbb{E}
\left[
\left\langle
\mathcal{C}_i,
\mathcal{C}_{i'}
\right\rangle
\right]
=
\left\langle
x_i,
x_{i'}
\right\rangle.
\label{eq:countsketch-inner-product}
\end{equation}
A contextual item $j$ shared by items $i$ and $i'$ contributes
\begin{equation}
\sigma(j)^2x_i[j]x_{i'}[j]
=
x_i[j]x_{i'}[j],
\end{equation}
whereas a cross term produced by unrelated items $j\ne j'$ contains
$\sigma(j)\sigma(j')$ and has zero expectation. Therefore, common local
contexts make consistent positive contributions, while unrelated hash
collisions do not systematically increase the raw-sketch similarity.

The sketch buckets may still contain large values caused by highly frequent
co-occurrences. We therefore apply a signed-log transformation:
\begin{equation}
z_i[\nu]
=
\operatorname{sign}
\left(
\mathcal{C}_i[\nu]
\right)
\log
\left(
1+\left|\mathcal{C}_i[\nu]\right|
\right).
\label{eq:cce-signed-log}
\end{equation}
This transformation preserves the sign of each aggregated bucket while
compressing its magnitude, thereby reducing the dominance of frequent
contextual items and large residual bucket values.

We then perform per-item $\ell_2$ normalization:
\begin{equation}
\bar{z}_i
=
\frac{z_i}
{\max\left(
\lVert z_i\rVert_2,
\vartheta_{\mathrm n}
\right)},
\label{eq:cce-normalization}
\end{equation}
where $\vartheta_{\mathrm n}>0$ is a numerical stability constant. A shared
fixed random projection maps the normalized sketch to the collaborative
embedding space:
\begin{equation}
\widetilde{e}_i^{\mathrm{col}}
=
\Lambda_{\mathrm{col}}\bar{z}_i,
\qquad
\Lambda_{\mathrm{col}}\in\mathbb{R}^{n_{\mathrm{col}}\times n_{\mathrm{bucket}}},
\label{eq:cce-projection}
\end{equation}
where $n_{\mathrm{col}}$ matches the semantic embedding dimension. The
projected representation is further normalized as
\begin{equation}
e_i^{\mathrm{col}}
=
\frac{
\widetilde{e}_i^{\mathrm{col}}
}{
\max\left(
\left\lVert
\widetilde{e}_i^{\mathrm{col}}
\right\rVert_2,
\vartheta_{\mathrm n}
\right)
}.
\label{eq:cce-embedding}
\end{equation}

\subsection{Confidence-Aware Weighting}

For each user $u$, let
$\mathcal R_u^+=(\jmath_{u,1},\ldots,\jmath_{u,L_u^+})$
denote the chronologically ordered high-quality positive-feedback training
sequence, where $\jmath_{u,p}$ is the item at position $p$ and $L_u^+$ is the
sequence length. Validation and test interactions are excluded from
$\mathcal R_u^+$.

Given the context-window radius $\omega$, user $u$ contributes contextual
information to item $i$ if $i$ occurs at some position $p$ in
$\mathcal R_u^+$ and
there exists another position $q$ within its context window. We define the
distinct-user support of item $i$ as
\begin{equation}
U_i
=
\sum_u
\mathbf{1}
\left[
\begin{gathered}
\exists\,p,q\in\{1,\ldots,L_u^+\}:\\
\jmath_{u,p}=i,\qquad 0<|p-q|\leq\omega
\end{gathered}
\right],
\label{eq:distinct-user-support}
\end{equation}
where $\mathbf{1}[\cdot]$ is an indicator function. For a given user $u$,
the indicator equals $1$ if the user's sequence provides at least one valid
contextual update for item $i$, and equals $0$ otherwise. Since each user
corresponds to only one binary indicator, repeated occurrences of item $i$
or multiple contextual updates from the same user do not increase $U_i$.
Therefore, $U_i$ counts the number of distinct users that contribute
collaborative evidence to item $i$, rather than the total number of
interactions.

Since distinct-user support is typically long-tailed, we first apply
logarithmic compression and then convert it into an item-level collaborative
confidence:
\begin{equation}
\gamma_i
=
\frac{\log(1+U_i)}
{\log(1+U_i)+\tau},
\qquad
\alpha_i
=
\alpha_{\mathrm{col}}\gamma_i,
\label{eq:appendix-caw}
\end{equation}
where $\tau>0$ controls the saturation speed of the confidence and
$\alpha_{\mathrm{col}}\in[0,1]$ specifies the maximum collaborative weight.
The logarithmic transformation compresses large differences among
high-support items and introduces diminishing returns as additional users
are observed. Consequently, items with sufficient behavioral support can
make greater use of collaborative information without allowing highly
popular items to dominate solely because of their interaction scale.

A smaller $U_i$ produces a smaller $\alpha_i$, downweighting potentially
unreliable collaborative evidence for interaction-sparse or cold-start
items. In particular, $U_i=0$ directly gives
$\gamma_i=\alpha_i=0$, so an item without valid collaborative support relies
entirely on its semantic representation.

Before fusion, the semantic representation is normalized as
\begin{equation}
\bar{e}_i^{\mathrm{sem}}
=
\frac{
e_i^{\mathrm{sem}}
}{
\max\left(
\lVert e_i^{\mathrm{sem}}\rVert_2,
\vartheta_{\mathrm n}
\right)
}.
\label{eq:semantic-normalization}
\end{equation}
The unified representation is then constructed through confidence-weighted
concatenation:
\begin{equation}
e_i^{\mathrm{uni}}
=
\left[
\sqrt{1-\alpha_i}\,
\bar{e}_i^{\mathrm{sem}};
\sqrt{\alpha_i}\,
e_i^{\mathrm{col}}
\right].
\label{eq:appendix-confidence-fusion}
\end{equation}
This concatenation retains the complementary structures of the semantic and
collaborative branches, while $\alpha_i$ controls their relative
contributions according to item-level behavioral support.

When both branches are nonzero and unit-normalized, the square-root
coefficients preserve the norm of the unified representation:
\begin{equation}
\begin{aligned}
\left\lVert e_i^{\mathrm{uni}}\right\rVert_2^2
&=
(1-\alpha_i)
\left\lVert\bar{e}_i^{\mathrm{sem}}\right\rVert_2^2
+
\alpha_i
\left\lVert e_i^{\mathrm{col}}\right\rVert_2^2\\
&=
(1-\alpha_i)+\alpha_i\\
&=1.
\end{aligned}
\label{eq:appendix-fusion-norm}
\end{equation}
Thus, under the stated nonzero, unit-normalized condition, CAW adjusts the
relative contribution of semantic and collaborative information without
introducing item-dependent variation in the overall representation scale.

The complete construction of the collaborative embeddings and their
confidence-aware fusion with the semantic representations is summarized in
Algorithm~\ref{alg:cce}.


\section{Appendix 5: Details of Reward-Consistent Calibration}

Due to confidentiality constraints, the full industrial reward schema cannot
be disclosed. We therefore report the retained feedback signals, fixed weights,
and exposure-only calibration procedure used in the experiments.

For user \(u\), let \(\mathcal E_u\) be the collection of complete slates
recorded in the training exposure log. Operationally, the ``candidate
rollouts'' in the main paper refer to evaluating logged candidate slates under
the policy, rather than generating new slates for reward assignment. The
alignment action set \(\mathcal A_u\subseteq\mathcal E_u\) is therefore formed
only from these logged exposure slates. For each \(a\in\mathcal A_u\), the
frozen reference policy and the current policy are used only to compute the
corresponding slate log-likelihoods \(\ell_0(a\mid u)\) and
\(\ell_\theta(a\mid u)\). Newly generated or counterfactual slates are not
assigned observed-feedback rewards and are excluded from reward calibration.

\begin{table}[H]
\centering
\footnotesize
\renewcommand{\arraystretch}{1.05}
\setlength{\tabcolsep}{2pt}
\begin{tabular}{@{}>{\raggedright\arraybackslash}p{0.28\columnwidth}
                    >{\raggedright\arraybackslash}p{0.66\columnwidth}@{}}
\toprule
\textbf{Component} & \textbf{Retained signals and fixed weights} \\
\midrule
Primary (positive) &
Effective view \(+0.10\); completion \(+0.15\); like \(+0.20\); share \(+0.15\). \\
Primary (negative) &
Immediate skip \(-0.15\); dislike \(-0.25\). \\
Auxiliary &
Diversity \(0.90\); novelty \(0.10\). \\
\bottomrule
\end{tabular}
\caption{Fixed reward components and weights used in alignment.}
\label{tab:reward-signals}
\end{table}

\paragraph{Primary Reward}
For an exposed item at position \(k\) of action \(a\), let
\(b_{u,a,k}^{\mathrm{ev}}\), \(b_{u,a,k}^{\mathrm{comp}}\),
\(b_{u,a,k}^{\mathrm{like}}\), \(b_{u,a,k}^{\mathrm{share}}\),
\(b_{u,a,k}^{\mathrm{skip}}\), and \(b_{u,a,k}^{\mathrm{dislike}}\) be binary
indicators for an effective view, completion, like, share, immediate skip,
and dislike, respectively. A zero means that the item was exposed but the
event did not occur. We use the following fixed slate reward:
\begin{equation}
\begin{aligned}
r_{u,a}^{\mathrm{pri}}
=\frac{1}{K}\sum_{k=1}^{K}\Bigl(
&0.10\,b_{u,a,k}^{\mathrm{ev}}
+0.15\,b_{u,a,k}^{\mathrm{comp}}\\
&+0.20\,b_{u,a,k}^{\mathrm{like}}
+0.15\,b_{u,a,k}^{\mathrm{share}}\\
&-0.15\,b_{u,a,k}^{\mathrm{skip}}
-0.25\,b_{u,a,k}^{\mathrm{dislike}}
\Bigr).
\end{aligned}
\label{eq:primary-reward-details}
\end{equation}
The six signals cover consumption quality, explicit positive actions, and
explicit negative actions. Effective view receives the smallest positive
weight; completion and share receive equal intermediate weights; and like
receives the largest positive weight. Dislike incurs a larger penalty than
immediate skip. The absolute weights sum to one, which keeps the per-position
scale easy to interpret.

\paragraph{Auxiliary Reward}
We retain only intra-slate diversity and novelty. Let \(\bar e_i\) be the
unit-normalized semantic representation of item \(i\), let \(c_i\) be its
exposure count in the training split, and define the Laplace-smoothed total
\(C=\sum_{j\in\mathcal I}(c_j+1)\). For the exposed item sequence
\((y_{u,a,1},\ldots,y_{u,a,K})\), the two scores are
\begin{equation}
\begin{aligned}
d_{u,a}
&=
\frac{2}{K(K-1)}
\sum_{1\leq p<q\leq K}
\frac{1-\bar e_{y_{u,a,p}}^\top\bar e_{y_{u,a,q}}}{2},\\
n_{u,a}
&=
\frac{1}{K}\sum_{k=1}^{K}
\frac{-\log\!\left((c_{y_{u,a,k}}+1)/C\right)}{\log C}.
\end{aligned}
\label{eq:auxiliary-components}
\end{equation}
Both terms lie in \([0,1]\): \(d_{u,a}\) is the mean pairwise cosine
distance, and \(n_{u,a}\) is larger for less frequently exposed items. The
auxiliary reward is
\begin{equation}
r_{u,a}^{\mathrm{aux}}
=0.90\,d_{u,a}+0.10\,n_{u,a}.
\label{eq:auxiliary-reward-details}
\end{equation}
These fixed weights are used throughout the alignment experiments. Diversity
receives the dominant weight because it directly measures redundancy at the
slate level; novelty is retained as a smaller secondary correction.

\paragraph{Relative Reward}
For \(o\in\{\mathrm{pri},\mathrm{aux}\}\), we compute statistics over the
same exposure-only action set:
\begin{equation}
\begin{aligned}
\operatorname{Mean}_u^o
&=
\frac{1}{|\mathcal A_u|}
\sum_{a\in\mathcal A_u}r_{u,a}^o,
\\
\left(\operatorname{Std}_u^o\right)^2
&=
\frac{1}{|\mathcal A_u|}
\sum_{a\in\mathcal A_u}
\left(r_{u,a}^o-\operatorname{Mean}_u^o\right)^2
.
\end{aligned}
\label{eq:reward-statistics}
\end{equation}
The relative reward score used in the main paper is
\begin{equation}
\delta_{u,a}^o
=
\frac{
r_{u,a}^o-\operatorname{Mean}_u^o
}{
\max(\operatorname{Std}_u^o,\varepsilon)
}.
\label{eq:relative-reward-score}
\end{equation}
A positive \(\delta_{u,a}^o\) indicates that the slate performs better than
the user's candidate average, while a negative value indicates worse relative
performance. Users with fewer than two logged slates are excluded from the
alignment batch. If a reward component is constant within \(\mathcal A_u\),
its standardized score is zero for every action and it contributes no
preference direction.

\raggedbottom
\section{Appendix 6: Hyperparameter Configuration}

Tables~\ref{tab:tusid-hyperparameters} and~\ref{tab:ogr-hyperparameters}
summarize the principal TUSID, OGR, and SPA configurations.
All offline results reported in the main paper and this appendix are means over
five independent runs with seeds 2025--2029.

\begin{table}[H]
\centering
\footnotesize
\renewcommand{\arraystretch}{1.04}
\setlength{\tabcolsep}{1mm}
\begin{tabular}{@{}>{\raggedright\arraybackslash}p{0.29\columnwidth}
                    >{\raggedright\arraybackslash}p{0.63\columnwidth}@{}}
\toprule
TUSID setting & Value \\
\midrule
Attributes & Brand; Category; B-tags; C-tags \\
Semantic fusion & Semantic dimensions 128; projection dimensions 512; 4 cross-attention layers; 8 heads \\
Regularization & $\beta_{\mathrm{res}}=0.001$ \\
CCE data & Positive-feedback training prefixes only \\
Sketch & $\omega=5$; $n_{\mathrm{bucket}}=256$; seed 2026 \\
Collaborative branch & Collaborative dimensions 128; $\tau=0.5$; $\alpha_{\mathrm{col}}=0.35$ \\
Quantization & RQ-KMeans; $D=4$; 1,024 codes per level \\
\bottomrule
\end{tabular}
\caption{Principal TUSID configurations.}
\label{tab:tusid-hyperparameters}
\end{table}

\begin{table}[H]
\centering
\footnotesize
\renewcommand{\arraystretch}{1.04}
\setlength{\tabcolsep}{1mm}
\begin{tabular}{@{}>{\raggedright\arraybackslash}p{0.29\columnwidth}
                    >{\raggedright\arraybackslash}p{0.63\columnwidth}@{}}
\toprule
OGR/SPA setting & Value \\
\midrule
Data and setup & Industrial; KuaiRec; history 128; $K=5$; AdamW; 5 runs (seeds 2025--2029); NVIDIA L20 \\
A/B testing time & 1 week \\
History encoder & 4 layers; 8 heads; hidden 512; FFN 2,048 \\
Planner & 2 layers; 8 heads; hidden 512; FFN 2,048 \\
SID decoder & 2 layers; 8 heads; hidden 512; FFN 2,048; beam width 20 \\
SID embedding & 512 dimensions; 1,024 code tokens plus one anomaly token; shared across 4 levels \\
Supervised decoding & $\alpha=0.3$ \\
Primary reward & Effective view $+0.10$; completion $+0.15$; like $+0.20$; share $+0.15$; immediate skip $-0.15$; dislike $-0.25$ \\
Auxiliary reward & Diversity $0.90$; novelty $0.10$ \\
SPA & $\varepsilon=10^{-6}$; $\epsilon=0.1$; $\gamma=0.05$; $\eta=0.1$; learning rate $10^{-5}$; batch 64; 3 epochs; planner and SID decoder trainable \\
\bottomrule
\end{tabular}
\caption{Principal OGR and SPA configurations.}
\label{tab:ogr-hyperparameters}
\end{table}
\FloatBarrier
\section{Appendix 7: Symbol Glossary}

This section consolidates notation used throughout the paper.
Tables~\ref{tab:main-symbols-problem}--\ref{tab:main-symbols-spa} list
main-paper symbols, while Tables~\ref{tab:appendix-symbols-rqcs}--%
\ref{tab:appendix-symbols-detail} list appendix-only notation; symbols reused
with the same meaning are not repeated.

\subsection{Symbols Used in the Main Paper}

\begin{table}[H]
\centering
\fontsize{8}{8.8}\selectfont
\captionsetup{skip=2pt}
\renewcommand{\arraystretch}{0.96}
\setlength{\tabcolsep}{1mm}
\begin{tabular}{@{}>{\raggedright\arraybackslash}p{0.29\columnwidth}
                    >{\raggedright\arraybackslash}p{0.63\columnwidth}@{}}
\toprule
Symbol & Meaning \\
\midrule
\(u\); \(i,j\) & User index; item indices. \\
\(t,p,q,T\) & History/context positions and history length. \\
\(k,m,K\) & Slate-position indices and slate size. \\
\(d,D\) & SID-depth index and total SID depth. \\
\(\mathcal H_u\) & Chronological interaction history of user \(u\). \\
\(\mathbf s_i,s_i^d\) & Hierarchical SID of item \(i\) and its depth-\(d\) token. \\
\(F_\theta,\theta\) & OGR mapping and its parameters. \\
\(\widehat{\mathcal S}_u,\widehat{\mathcal Y}_u\) & Generated SID slate and mapped item slate. \\
\(\mathrm{Map}(\cdot)\) & SID-to-item lookup. \\
\(\mathcal Y_u,\mathcal W_u,y_k,w_k\) & Exposure slate, feedback sequence, item, and feedback at position \(k\). \\
\(e_i^{\mathrm M}\) & Fine-grained multimodal representation. \\
\(\mathrm a,e_i^{\mathrm a}\) & Attribute-source identifier and representation, with
\(\mathrm a\in\{\mathrm b,\mathrm c,\mathrm{Bt},\mathrm{Ct}\}\). \\
\(\mathcal E_i^{\mathrm{attr}}\) & Stack of the four attribute representations. \\
\(Q_i,K_i,V_i\) & Cross-attention query, keys, and values. \\
\(h_i^{\mathrm{attr}},g_i,e_i^{\mathrm{sem}}\) & Attribute aggregate, residual gate, and fused semantic representation. \\
\(\mathcal L_{\mathrm{rec}},\mathcal L_{\mathrm{fusion}}\) & Recommendation and Semantic-Fusion objectives. \\
\(\mathcal B,\beta_{\mathrm{res}}\) & Fusion training batch and residual-regularization coefficient. \\
\bottomrule
\end{tabular}
\caption{Main-paper notation for Problem Formulation and Semantic Fusion.}
\label{tab:main-symbols-problem}
\end{table}

\newpage
\begin{table}[H]
\centering
\footnotesize
\renewcommand{\arraystretch}{1.02}
\setlength{\tabcolsep}{1mm}
\begin{tabular}{@{}>{\raggedright\arraybackslash}p{0.29\columnwidth}
                    >{\raggedright\arraybackslash}p{0.63\columnwidth}@{}}
\toprule
Symbol & Meaning \\
\midrule
\(\omega\) & Local co-occurrence window radius. \\
\(\mathcal C_i,h,\sigma\) & Collaborative sketch, bucket hash, and sign hash. \\
\(e_i^{\mathrm{col}}\) & Dense collaborative representation. \\
\(U_i\) & Number of distinct users supporting item \(i\). \\
\(\gamma_i,\tau\) & Collaborative confidence and smoothing coefficient. \\
\(\alpha_i,\alpha_{\mathrm{col}}\) & Item-specific and maximum collaborative weights. \\
\(e_i^{\mathrm{uni}}\) & Unified semantic--collaborative representation. \\
\(\mathcal E\) & Shared SID embedding table. \\
\(\mathrm{Enc}_{\mathrm{hist}},p_t^{\mathrm h},\mathcal Z^{\mathrm h}\) & History encoder, temporal embedding, and encoded history. \\
\(g_m\) & Aggregated target SID embedding at slate position \(m\). \\
\(\mathcal Q^{\mathrm{tr}},\mathrm{BOS}\) & Teacher-forced planner input and beginning-of-sequence token. \\
\(\mathcal P,p_m,\mathrm{Dec}_{\mathrm{plan}}\) & Planned preference sequence, position-\(m\) representation, and planner. \\
\(\mathbf s_m,s_m^{<d},\mathrm{Dec}_{\mathrm{sid}}\) & Decoded SID, its prefix, and position-wise SID decoder. \\
\(\mathcal Y_u^{\mathrm{eps}},\mathcal Y_u^{\mathrm{fb}}\) & Exposure-order and feedback-order supervision slates. \\
\(o\) & Branch identifier: \(\mathrm{eps}/\mathrm{fb}\) for supervision and
\(\mathrm{pri}/\mathrm{aux}\) for reward calibration. \\
\(\mathcal U\) & Set of training users. \\
\(s_{u,m}^{o,d},\pi_{u,m}^{o,d}\) & Target SID token and predicted token distribution. \\
\(\mathcal L_{\mathrm{sid}}^o,\mathcal L_{\mathrm{sup}},\alpha\) & Branch SID loss, supervised objective, and feedback-branch weight. \\
\bottomrule
\end{tabular}
\caption{Main-paper notation for Collaborative Injection and GL2P.}
\label{tab:main-symbols-gl2p}
\end{table}

\newpage
\begin{table}[H]
\centering
\footnotesize
\captionsetup{skip=2pt}
\renewcommand{\arraystretch}{0.98}
\setlength{\tabcolsep}{1mm}
\begin{tabular}{@{}>{\raggedright\arraybackslash}p{0.27\columnwidth}
                    >{\raggedright\arraybackslash}p{0.65\columnwidth}@{}}
\toprule
Symbol & Meaning \\
\midrule
\(\theta_0;\ell_0,\ell_\theta\) & Frozen reference-policy parameters; reference and current slate log-likelihoods. \\
\(\mathcal A_u,a;\mathbf s_{u,k}\) & Candidate slate-action set and action; candidate SID at position \(k\). \\
\(r_{\mathrm{pri}},r_{\mathrm{aux}};\delta_{u,a}^o\) & Primary and auxiliary slate rewards; within-user standardized reward component. \\
\(\begin{gathered}\kappa_{u,a},\Delta_{u,a};\\ \varepsilon,\rho_{u,a}\end{gathered}\) & Agreement gate and calibrated preference signal; numerical floor and likelihood ratio. \\
\(\mathcal L_{\mathrm{pol}},N_A,\epsilon\) & Clipped policy loss, retained action count, and clipping radius. \\
\(q_0,q_\theta;\mathcal L_{\mathrm{KL}},N_U\) & Reference/current action distributions; candidate-distribution KL loss and retained-user count. \\
\(\gamma,\eta,\mathcal L_{\mathrm{SPA}}\) & KL weight, supervised-replay weight, and SPA objective. \\
HR, Recall, NDCG; ICR, CUR & Ranking metrics at the stated cutoff; SID uniqueness and codebook coverage. \\
Min.\ PPL, Top-1 Load; V-measure, SC & Minimum assignment perplexity and maximum code load; category consistency and semantic cohesion. \\
\(\begin{gathered}\mathrm{Work}_{\mathrm{serial}},\\
\mathrm{Work}_{\mathrm{OGR}}\end{gathered}\) & Total serial and OGR decoding work. \\
\(\begin{gathered}\mathrm{cost}_t^{\mathrm{serial}},\\
\mathrm{cost}_m^{\mathrm{plan}},\\
\mathrm{cost}_d^{\mathrm{sid}}\end{gathered}\) & Per-stage serial, planning, and SID-decoding costs. \\
\(\begin{gathered}\mathrm{Depth}_{\mathrm{serial}},\\
\mathrm{Depth}_{\mathrm{OGR}}\end{gathered}\) & Sequential dependency depths. \\
\bottomrule
\end{tabular}
\caption{Main-paper notation for SPA, Evaluation, and Efficiency Analysis.}
\label{tab:main-symbols-spa}
\end{table}
\FloatBarrier

\subsection{Symbols Introduced Only in the Appendix}

\begin{table}[H]
\centering
\footnotesize
\captionsetup{skip=2pt}
\renewcommand{\arraystretch}{0.98}
\setlength{\tabcolsep}{1mm}
\begin{tabular}{@{}>{\raggedright\arraybackslash}p{0.31\columnwidth}
                    >{\raggedright\arraybackslash}p{0.61\columnwidth}@{}}
\toprule
Symbol & Meaning \\
\midrule
\(N_{\mathrm{item}};n_d^{\mathrm{cb}},n\) & Number of quantized items; depth-\(d\) codebook size and centroid index. \\
\(\Xi^{(d)},\xi_n^{(d)};\zeta_i^{(d)},\widehat e_i^{\mathrm{rq}}\) & Depth-\(d\) codebook and centroid; pre-quantization residual and reconstruction. \\
\(\begin{gathered}n_{\mathrm{vocab}};\\ n_{\mathrm{row}},n_{\mathrm{bucket}};v,\nu\end{gathered}\) & Coordinate-universe size; CountSketch row/bucket counts and indices. \\
\(x;\mathbf G_{\mathrm{cs}}\) & Classical CountSketch frequency vector and table. \\
\(\mathfrak h_v,\varsigma_v\) & Row-specific bucket and sign hashes. \\
\(\upsilon;\widehat x_j^{(v)}\) & Stream-update magnitude and row-wise estimate; \(j\) retains its main-paper meaning. \\
\bottomrule
\end{tabular}
\caption{Appendix-only notation for RQ-KMeans and CountSketch preliminaries.}
\label{tab:appendix-symbols-rqcs}
\end{table}

\begin{table}[H]
\centering
\footnotesize
\renewcommand{\arraystretch}{1.02}
\setlength{\tabcolsep}{1mm}
\begin{tabular}{@{}>{\raggedright\arraybackslash}p{0.31\columnwidth}
                    >{\raggedright\arraybackslash}p{0.61\columnwidth}@{}}
\toprule
Symbol & Meaning \\
\midrule
\(\mathcal R_u^+,\jmath_{u,p},L_u^+\) & Positive-feedback training prefix, its item, and its length. \\
\(x_i\) & Uncompressed local-context vector. \\
\(\mathcal I_u^{\mathrm{ctx}}\) & Items receiving a valid context update from user \(u\). \\
\(z_i,\bar z_i\) & Signed-log sketch and normalized form. \\
\(\Lambda_{\mathrm{col}},n_{\mathrm{col}}\) & Fixed collaborative projection and output dimension. \\
\(\vartheta_{\mathrm n}\) & Numerical floor used in appendix normalization. \\
\(\widetilde e_i^{\mathrm{col}},\bar e_i^{\mathrm{sem}}\) & Projected collaborative and normalized semantic embeddings. \\
\(\mathcal E_u\) & Complete slates recorded in the training exposure log. \\
\(\begin{gathered}
b_{u,a,k}^{\mathrm{ev}},b_{u,a,k}^{\mathrm{comp}},\\
b_{u,a,k}^{\mathrm{like}},b_{u,a,k}^{\mathrm{share}},\\
b_{u,a,k}^{\mathrm{skip}},b_{u,a,k}^{\mathrm{dislike}}
\end{gathered}\) & Binary indicators for the six retained feedback signals. \\
\(\bar e_i,c_i,C\) & Unit-normalized semantic representation, training exposure count, and Laplace-smoothed total. \\
\(y_{u,a,k},d_{u,a},n_{u,a}\) & Exposed item, intra-slate diversity score, and novelty score. \\
\(\operatorname{Mean}_u^o,\operatorname{Std}_u^o\) & Candidate-set reward mean and standard deviation. \\
\bottomrule
\end{tabular}
\caption{Appendix-only notation for CCE and detailed rewards.}
\label{tab:appendix-symbols-detail}
\end{table}

\subsection{Notation Conventions}

Symbols inherited from the main paper retain exactly the same meaning in the
appendix. In particular, \(\alpha\) and \(\gamma\) without item subscripts are
training-objective weights, whereas \(\alpha_i\), \(\alpha_{\mathrm{col}}\),
and \(\gamma_i\) belong to confidence-aware collaborative fusion. The reward
floor \(\varepsilon\) and policy clipping radius \(\epsilon\) are also distinct.
Appendix-only symbols use distinct names or typefaces and are summarized in
Tables~\ref{tab:appendix-symbols-rqcs}--\ref{tab:appendix-symbols-detail}.
\FloatBarrier

\end{document}